# Investigation of production of neutral Higgs boson and two charged charginos from electron-positron annihilation via different propagators

Sara Abdelrady Hassan[1], Asmaa.A. A[2], Sherif Yehia[3], M.M. Ahmed[3] and Mohammed Said Mohammed Abu-Elmagd[1]

**[1]**Department of Engineering Mathematics and Physics, Higher Institute of Engineering, El-Shorouk Academy, El-Shorouk City, Egypt

**[2]**Egyptian Organization for Standardization and Quality, Cairo, Egypt

**[3]**Department of Physics, Faculty of Science, Helwan University, Cairo, Egypt

* Fax: +2(02)26000039 E-mail: m.said@sha.edu.eg

## Abstract

In the current work we investigated the production of neutral Higgs boson ($H_\ell^\circ$) and two charged Charginos ($\tilde{\chi}_i^+, \tilde{\chi}_j^-$) owing to electron-positron annihilation via different propagators for the process $e^-(p_1) + e^+(p_3) \rightarrow \tilde{\chi}_i^+(p_2) + \tilde{\chi}_j^-(p_4) + H_\ell^\circ(p_5)$ and in the Minimal Supersymmetric Standard Model (MSSM), the cross sections for this interaction were estimated. Five groups of 180 probabilities from Feynman diagrams are taken by different propagators. Group (I) when $h^o$ and $Z^o$ bosons are propagators, Group (II) when $Z^o$ and $h^o$ bosons are propagators, Group (III) when $h^o$ and $h^o$ (Lightest Higgs boson) bosons are propagators, Group (IV) $Z^o$ and $Z^o$ bosons are propagators and Group (V) $\tilde{\chi}^o$ and $Z^o$ bosons are propagators where $i = j = 1,2$ and $\ell = 1,2,3$.

We calculated the process's production cross-sections as a function of mass center energy, and determined the best cross-section based on all considerations of the (MSSM), the process's mechanisms can be identified as:

$$e^-(P_1) + e^+(P_3) \rightarrow Z^o(P_1 + P_3) \rightarrow Z^o(P_2 + P_4) \rightarrow \tilde{\chi}_i^+(p_2) + \tilde{\chi}_j^-(p_4) \text{ in group IV}$$

$$e^-(P_1) + e^+(P_3) \rightarrow Z^o(P_1 + P_3) \rightarrow h^o(P_2 + P_4) \rightarrow \tilde{\chi}_i^+(p_2) + \tilde{\chi}_j^-(p_4) \text{ in group II.}$$

$$e^-(P_1) + e^+(P_3) \rightarrow h^o(P_1 + P_3) \rightarrow h^o(P_2 + P_4) \rightarrow \tilde{\chi}_i^+(p_2) + \tilde{\chi}_j^-(p_4) \text{ in group III}$$

$$e^-(P_1) + e^+(P_3) \rightarrow \tilde{\chi}^o(P_1 + P_3) \rightarrow Z^o(P_2 + P_4) \rightarrow \tilde{\chi}_i^+(p_2) + \tilde{\chi}_j^-(p_4) \text{ in group V}$$

$$e^-(P_1) + e^+(P_3) \rightarrow h^o(P_1 + P_3) \rightarrow Z^o(P_2 + P_4) \rightarrow \tilde{\chi}_i^+(p_2) + \tilde{\chi}_j^-(p_4) \text{ in group I.}$$

At S interval (1600- 3500) Gev, the best value of σ is ( $7.3934 \times 10^{-4}$ ) Pb in-group (IV). When masses of Charginos are $m_{\tilde{\chi}_i^-} = 900$ GeV, $m_{\tilde{\chi}_j^+} = 700$ GeV and mass of neutral Higgs boson is $m_{H_\ell^\circ} = 140$ GeV

**Keywords:** Higgs boson, Chargino and Neutralino

## 1. Introduction

The Supersymmetry Standard Model (SUSY) [1–7], suggests adding a new symmetry to particle physics' Standard Model (SM), as well as a symmetry between bosons and fermions, and anticipates the presence of potential partners for each Standard Model (SM) particle. This provides resolve for the hierarchy dilemma [7-12] and a nominee for dark matter in the form of the lightest supersymmetric particle (LSP), which will be static in the situation of conserved R-parity [13].

The SM's minimal supersymmetric extension (MSSM) [14, 15], The bino, the winos, and the Higgsino are the superpartners of the U(1)$_Y$ and SU(2)$_L$ gauge fields, as well as the Higgs field. The mass terms for the bino, wino, and Higgsino states are M$_1$,M$_2$, and, μ respectively. Since they not carry color charge, they can only be produced through electroweak interactions or the decay of colored superpartners. Because electroweak processes have smaller cross sections, the masses of these objects are observationally less limited than the masses of colored SUSY particles. According to the mass spectrum. Through mixing of the superpartners, chargino ($\tilde{\chi}_{1,2}^{\pm}$) and neutralino ($\tilde{\chi}_{1,2,3,4}^{0}$ ) mass eigenstates are created. These are known as electroweakinos, and the subscripts imply increasing electroweakino mass. If the $\tilde{\chi}_{1,2,3,4}^{0}$ is stable, for example as the lightest supersymmetric particle (LSP) and R-parity conservation is postulated, it is a viable dark-matter candidate [16, 17].

This paper calculates the cross sections (σ) as a function of center of mass energy a search for direct production of neutral Higgs boson and two charged charginos from electron-positron annihilation via different propagators for the process $e^-(p_1) + e^+(p_3) \rightarrow \tilde{\chi}_i^+(p_2) + \tilde{\chi}_j^-(p_4) + H_\ell^{\circ}(p_5)$

## 1.1 The Cross-section Scattering

In physics, the significance of cross section is an indicator of the probability that a particular process will occur when a particular type of radiant excitation encounters a highly concentrated phenomenon. The Rutherford cross-section, for example, is an indicator of the chance of an alpha particle being diverted by a specific direction throughout an interaction with an atomic nucleus. σ (sigma) cross section and is measured in term of area, specifically barns. In some ways, it can be compared to the size of the object that the excitation must strike throughout order for the process to take place.

We have learned a lot about nuclear and atomic physics through scattering experiments, such as the discovery of subatomic particles (such as quarks). Scattering phenomena, such as neutron, electron, and x-ray scattering, are used to investigate solid state systems in low energy physics. As a main overview, it is therefore essential in any advanced quantum mechanics course.

If the radiation is thought to be made up of quanta, The quantity of incident particles hitting the target's surface per unit time per unit area is the flux, and the cross-section measures the scattering rate per unit incident radiation flux. Calculating scattering cross-sections for long-wavelength electromagnetic radiation means dividing the power of the scattered wave by the intensity of the incident wave. A cross-section represents an area in dimensions, with its unit is barn, which has an area of $10^{-28}$ m$^2$. Instead of a true geometric cross-sectional area, a scattering cross-section can be interpreted as an effective area proportional to the probability of interaction between the radiation and the target.

A differential cross section is the differential limit of a function of some final-state parameter, such as particle angle or energy. A total cross section or integrated total cross section is a cross section that has been integrated over all scattering angles. In a real scattering experiment, the different rates of scattering to different angles provide information about the scatterer. Detectors are positioned at various angles $(\theta, \phi)$. The standard form for an infinitely small solid angle is $d\Omega = sin\theta d\theta d\phi$. The total solid angle (all probable scatterings) is $\int d\Omega = 4\pi$ the area of a unit radius sphere.

The differential cross section, $d\sigma/d\Omega$, is the part of the total number of scattered particles which emerge in the solid angle $d\Omega$, so the rate of particle scattering to this detector is $nd\sigma/d\Omega$, with n defined above as the beam intensity. By integrating over all solid angles, we can obtain the total cross section from the differential.

$$\sigma = \int \frac{d\sigma}{d\Omega} d\Omega = \int_0^{2\pi} d\varphi \int_0^{\pi} d\theta sin\theta \frac{d\sigma}{d\Omega}$$

The cross section is sensitive to the energy of the incoming particles.

## 1.2 Properties of Chargino:

Chargino are composed of Winos ($W^+, W^-$) and Higgsinos ($H^+, H^-$)[17,18]. In nature, neutralino dark matter is experimentally investigated indirect through the use of $\gamma$ ray and neutrino telescopes or directly through the utilizing laboratory experiment like those of Cryogenic dark matter search (CDMS) [19, 20]. Heavier neutralinos usually disintegrate to lighter neutralinos via a neutral Z boson or a via charged W boson to a light chargino. [21]. It is created in pairs through s-channel $\gamma/Z$ exchange [22, 23]. The lightest neutralino $\tilde{\chi}_1^{\circ}$ is heavier than the lightest chargino $\tilde{\chi}_1^{\pm}$.

The mean lifetime $(\tau_{\tilde{\chi}_1^{\pm}})$ of $\tilde{\chi}_1^{\pm}$ is expressed in the form of $\Delta m_{\tilde{\chi}_1}$, which is customarily in the range of a nanosecond. Charginos have a lifetime ranging from 0.1 to 10 ns [24]. The charginos disintegrate to the lightest neutralino $\tilde{\chi}_1^{\circ}$, that is believed to be stable, and a two fermions (f) consisting of quarks and antiquarks or leptons and neutrinos [25]. The lightest chargino with a mass larger than 103.5 GeV [26].There are three variables or soft terms in the chargino mass matrix ($M_2$, $\mu$ and $\tan\beta$) and the neutralino mass matrix has four soft terms ( $M_1$, $M_2$, $\mu$ and $\tan\beta$).

## 1.3 Higgs boson:

The Standard Model (SM) of elementary particles describes strong and electroweak interactions between quarks and leptons by exchanging force carriers, such as photons for electromagnetic interactions, W and Z bosons for weak interactions, and gluons for strong interactions. Quarks and leptons serve as the fundamental components of matter in the SM. The electroweak hypothesis unifies the electromagnetic and weak interactions. The SM's predictions have been amply verified because they are remarkably compatible with the majority of accurate measurements up to the energies now available, but it is still unclear how the W and Z gauge bosons pick up mass while the photon stays massless. It was postulated almost 50 years ago that the introduction of a scalar field may lead to spontaneous symmetry breaking in gauge theories. The W and Z masses are produced as a result of applying this method to the electroweak theory through a complex scalar doublet field, and the SM Higgs boson's existence is predicted (H).Through the Yukawa interaction, the scalar field also provides mass to the fundamental fermions [27, 28].

### 1.3.1   Higgs Field

In the Standard Model, no particle has a mass when it first appears. This may be the case for photons, however the W and Z have mass that is close to 100 times greater than that of a proton. A solution was offered by Peter Higgs to this issue. The W and Z are given mass by the Brout-Englert-Higgs process, which interacts with an ethereal field that is now known as the Higgs field. The Higgs field was initially zero after the big bang, but when the cosmos cooled and the temperature dipped to a crucial point, it developed spontaneously, giving any particle interacting with it a mass. A particle becomes heavier as it interacts more with this field. The photon and other particles that do not interact with it have no mass at all. The Higgs boson is a particle that is connected to the Higgs field, like all other fundamental fields.

### 1.3.2   Properties of Higgs boson

The Higgs boson has no spin, has zero electric and color charge and it is also its own antiparticle.

### 1.3.3   The Higgs boson production and decay

The Higgs boson can be produced at the Large Hadron Collider (LHC) through gluon-gluon fusion (ggF) 87% of the time, accompanied by vector boson fusion (VBF, 7%), associated top-antitop production (ttH, 1%), which is known as the top fusion process because it includes two colliding gluons, where every decay into a heavy quark-antiquark pair. Each pair's quark and antiquark can then merge to create a Higgs particle.[29]

## 1.4  Two-Higgs-doublet model

Following the breakthrough of the Standard Model (SM) Brout-Englert-Higgs boson at the Large Hadron Collider (LHC), the concern of whether there are more particles within experimental reach remains open.

One straightforward possibility is that a second Higgs doublet has the same quantum numbers as the SM Higgs [30]. The Two-Higgs-Doublet Model (2HDM) is the most basic evolution of the Standard Model (SM), containing one extra scalar doublet with more physical neutral and charged Higgs fields.

There are five physical scalar states with the second Higgs doublet, including the CP even neutral Higgs bosons h and H (where H is heavier than h), the CP odd pseudoscalar A, and two charged Higgs bosons $H^{\pm}$. The Higgs boson detected is measured to be CP even. Six physical parameters, including four Higgs masses

$(m_h, m_H, m_A, and\ m_{H^\pm})$, the ratio of the two vacuum expectation values ($\tan \beta$), and the mixing angle ($\alpha$) that diagonalizes the mass matrix of the neutral CP even Higgses, can be used to characterise such a model. The mass of the Higgs particle and its vacuum expectation value are the only two parameters used by the SM.

$$m_{H^\pm}^2 = \lambda_4(v_1^2 + v_2^2)$$

$$m_A = \lambda_6(v_1^2 + v_2^2)$$

$$m_{H,h}^2 = \frac{1}{2}[a + C \pm D]$$

Were

$$a = 4v_1^2(\lambda_2 + \lambda_3) + v_2^2\lambda_5 \quad ; \quad B = (4\lambda_3 + \lambda_5)v_1 v_2$$

$$C = 4v_2^2(\lambda_2 + \lambda_3) + v_1^2\lambda_5 \quad ; \quad D = \sqrt{(A - C)^2 + 4B^2}$$

$\lambda$ : wavelength $\qquad ; \quad v$ : velocity

The Higgs potential that is composed of quadratic terms and quadratic interaction terms, governs the Higgs characteristics of the MSSM. Supersymmetric gauge couplings directly impact the strength of the interaction terms. The Higgs spectrum, an angle $\alpha$ (which represents the degree of mixing of the original $Y = \pm 1$ Higgs doublet states in the physical CP-even scalars), and the Higgs boson couplings are all determined by $tan\ \beta$ and one Higgs mass ($m_A$).

## 2. Rules of Calculation Cross sections in (Pb):

Initial states have momenta $p_1, p_3$ and their masses $m_1, m_3$, while three-body final states have momenta $p_2, p_4, p_5$ and their masses $m_2, m_4, m_5$.

$$p_1 + p_3 = p_2 + p_4 + p_5 \qquad (1)$$

$$s = \sigma + p_5 \qquad (2)$$

The cross section ($\sigma$) for the process $e^-(p_1) + e^+(p_3) \rightarrow \tilde{\chi}_i^+(p_2) + \tilde{\chi}_j^-(p_4) + H_\ell^*(p_5)$ can be expressed in writing as

$$\sigma = \int \pi^2 |M|^2 \frac{dx\ dy\ d\sigma^2}{\Lambda(S, m_1, m_3)\Lambda(S, \sigma, m_5)} \qquad (3)$$

Where M is the matrix element, by using Feynman rules this allows us to write the M-matrix as well as the trace thermos needed to compute the square matrix ( $|M|^2$ ), with the integration carried out by a straightforward approximation produced by an enhanced Weizsacker-Williamson approach [31, 32]. Where:

$$\Lambda(x, y, z) = [x^4 + y^4 + z^4 - 2x^2 y^2 - 2x^2 z^2 - 2y^2 z^2]^{1/2} \qquad (4)$$

Then, the integration limit and integration simplification which done by using the Mathematica application are.

$$x_\pm = \frac{1}{4S^2}[(S^2 + m_1^2 - m_3^2)(S^2 - \sigma^2 + m_5^2) \pm \Lambda(S, m_1, m_3)\Lambda(S, \sigma, m_5)] \qquad (5)$$

$$y_\pm = \frac{1}{4\sigma^2}[(\sigma^2 + m_2^2 - m_4^2)(S^2 - \sigma^2 + m_5^2) \pm \Lambda(\sigma, m_2, m_4)\Lambda(S, \sigma, m_5)] \qquad (6)$$

$$(m_2 + m_4)^2 \leq \sigma^2 \leq (S^2 - m_5^2)^2 \qquad (7)$$

After calculating the cross sections, the results subsequently charted and reported in tables. We employ the vector-boson masses given by [33] in every one of the computations.

## 2.1 Constants of the reaction.

$M_W = 80$ GeV

$M_Z = 90$ GeV

## 2.2 New variables of the reaction.

$M_{H_\ell^\circ}$, $\ell = 1,2,3$ where ($\ell_1 = M_{h^0}$), ($\ell_2 = M_{A^0}$), ($\ell_3 = M_{H^0}$)

$M_{h^0} = 125$ GeV (mass of $h^0$ propagator)

$M_{A^0} = 135$ GeV (mass of $A^0$ propagator)

$M_{H^0} = 140$ GeV (mass of $H^0$ propagator)

$m_{\tilde{\chi}_j^-} = (800 , 900)$GeV , $m_{\tilde{\chi}_i^+} = (600 , 700)$ GeV (i, j=1, 2)

# 3. Classified the reaction $e^-(p_1) + e^+(p_3) \rightarrow \tilde{\chi}_i^+(p_2) + \tilde{\chi}_j^-(p_4) + H_\ell^\circ(p_5)$ according to propagators.

There are five groups from Feynman diagrams classified according to the propagators within (MSSM) models

## 3.1  Group (I) via  $h^0$ and $Z^0$ propagators

### 3.1.1   Feynman Diagram for Group (I)

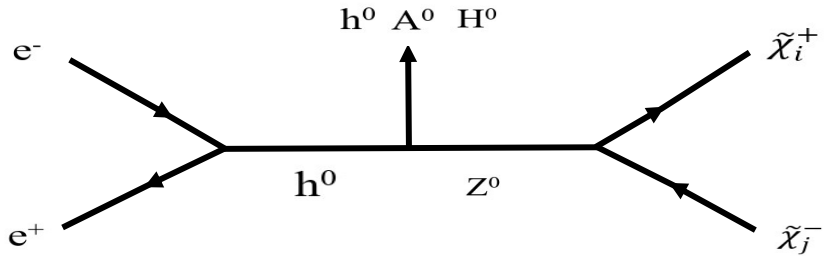

**Fig.1:** Feynman diagrams for the process $e^-(p_1) + e^+(p_3) \rightarrow \tilde{\chi}_i^+(p_2) + \tilde{\chi}_j^-(p_4) + H_\ell^\circ(p_5)$ via $h^0$ and $Z^0$ propagators,

### 3.1.2    Matrix element and Probability of reaction for Group (I)

**There should be 36 possibilities from (1– 36) and the Matrix element is:**

$M_{(1-36)} = (-i(F_L + \gamma_5 F_L^{'})U_{e^-}(P_1)\bar{V}_{e^+}(P_3)(iG_m(p\text{-}q)^\mu)(\sigma^2 - m_Z^2)^{-1}(S^2 - m_H^2)^{-1}(e\frac{\cos^2\theta_w - \sin^2\theta_w}{2\sin\theta_w\cos\theta_w}(P_2 - P_1))V_{\tilde{\chi}^+}(P_2)\bar{U}_{\tilde{\chi}^-}(P_4)$ [34]

Where:

$g$: The gauge coupling constants of $SU(2)_L$

$g = \frac{e}{Sin\,\theta_w} = 0.637132$

$m_e$: the mass of electron

$M_Z$: the mass of Z boson

$M_W = 80$ GeV

$M_Z = 90$ GeV

$M_{h^0} = 125$ GeV (mass of $h^0$ propagator, The lightest neutral Higgs boson)

$M_{A^0} = 135$ GeV (mass of $A^0$ propagetor)

$M_{H^0} = 140$ GeV (mass of $H^0$ propagetor, The heaviest neutral Higgs boson)

As, $m_{H_\ell^\circ}$, $\ell = 1,2,3$ as $\ell = (1=h^0)$, $(2=A^0)$, $(3=H^0)$

$m_{\tilde{\chi}_j^-} = (800\ ,\,900)$ GeV , $m_{\tilde{\chi}_i^+} = (600\ ,\,700)$ GeV  $(i, j=1,\ 2)$

$F_L = -g\left(\frac{m_e Cos\alpha}{2M_W\, Cos\,\beta}\right)$, $F_L^` = g\left(\frac{m_e Cos\alpha}{2M_W\, Cos\,\beta}\right)$ [35]

$G_m = ig\left(\frac{Sin(\beta-\alpha)}{2\,Cos\,\theta_w}\right)$

### 3.1.3   Cross Section Calculations in (Pb) for Group (I):

In this section we compute the cross sections as a function of center of mass energy for the Feynman diagram in fig. (1) using Feynman rules, equation (3), and the Mathematica software (1). the results shown in figs.2 (a-c) by varying the mass of charginos ($m_{\tilde{\chi}_i^+}$, $m_{\tilde{\chi}_i^-}$) at different masses of neutral Higgs boson

($M_{h^0}$, $M_{A^0}$, $M_{H^0}$)

for the process $e^-(p_1) + e^+(p_3) \rightarrow \tilde{\chi}_i^+(p_2) + \tilde{\chi}_j^-(p_4) + H_\ell^\circ(p_5)$

$m_{\left(\tilde{\chi}_i^+\ ,\,\tilde{\chi}_j^-\right)} \rightarrow m_{(600\ ,\,800)} \rightarrow m_{11}$ (Blue)

$m_{\left(\tilde{\chi}_i^+\ ,\,\tilde{\chi}_j^-\right)} \rightarrow m_{(600\ ,\,900)} \rightarrow m_{12}$  (Red)

$m_{\left(\tilde{\chi}_i^+\ ,\,\tilde{\chi}_j^-\right)} \rightarrow m_{(700\ ,\,800)} \rightarrow m_{21}$ (Green)

$m_{\left(\tilde{\chi}_i^+\ ,\,\tilde{\chi}_j^-\right)} \rightarrow m_{(700\ ,\,900)} \rightarrow m_{22}$ (Pink)

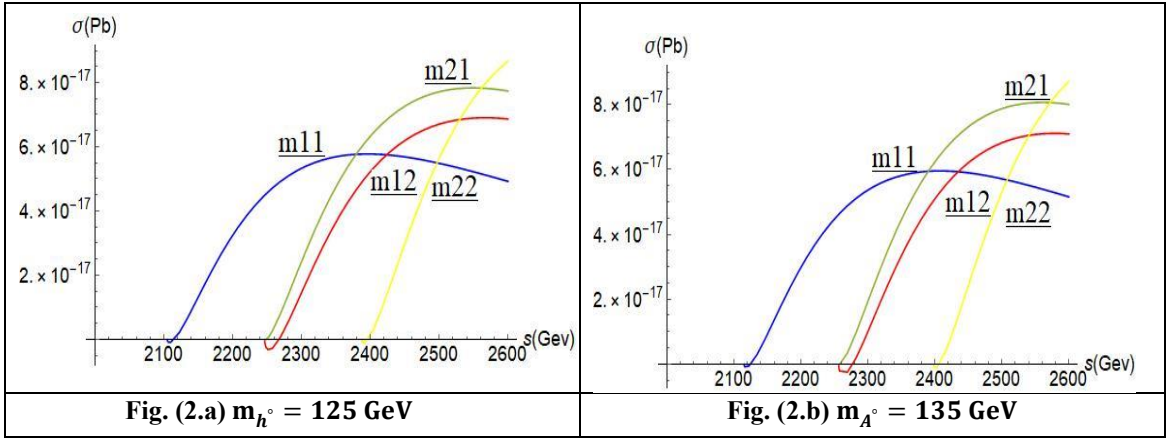

**Fig. (2.a) m_{h°} = 125 GeV**

**Fig. (2.b) m_{A°} = 135 GeV**

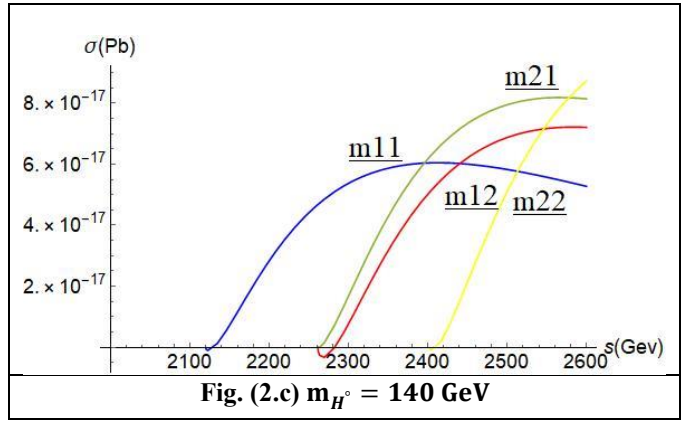

**Fig. (2.c) m_{H°} = 140 GeV**

**Fig.2:** The cross sections for the process $e^-(p_1) + e^+(p_3) \rightarrow \tilde{\chi}_i^+(p_2) + \tilde{\chi}_j^-(p_4) + H_\ell^°(p_5)$ as a function of center of mass energy via $h^o$ and $Z^o$ bosons propagators by interchanging the mass of Charginos $(m_{\tilde{\chi}_i^+}, m_{\tilde{\chi}_j^-})$ at different mass of neutral Higgs boson ( $M_{h^o}, M_{A^o}, M_{H^o}$ )

### 3.1.4 Comparing Results and Discussion for Group (I):

**Table (1):** Comparing between the cross sections for the process $e^-(P_1) + e^+(P_3) \rightarrow h^o(P_1 + P_3) \rightarrow Z^o(P_2 + P_4) \rightarrow \tilde{\chi}_i^+(p_2) + \tilde{\chi}_j^-(p_4)$ via $h^o$ and $Z^o$ propagators by interchanging the masses of Charginos $(m_{\tilde{\chi}_i^+}, m_{\tilde{\chi}_j^-})$ at different mass of resultant neutral Higgs boson ( $M_{h^o}, M_{A^o}, M_{H^o}$ )

| $e^-(\mathbf{P_1}) + e^+(\mathbf{P_3}) \rightarrow h^o(\mathbf{P_1 + P_3}) \rightarrow Z^o(\mathbf{P_2 + P_4}) \rightarrow \tilde{\chi}_i^+(\mathbf{p_2}) + \tilde{\chi}_j^-(\mathbf{p_4})$ | | | | | | |
|---|---|---|---|---|---|---|
| $\mathbf{m_{\tilde{\chi}_i^+}, m_{\tilde{\chi}_j^-}}$ **i,j = 1,2** | **Resultant $m_{h^o}$=125** | | **Resultant $m_{A^o}$=135** | | **Resultant $m_{H^o}$=140** | |
| | **Fig. (2.a)** | | **Fig. (2.b)** | | **Fig. (2.c)** | |
| | **S(Gev)** | **σ(Pb)** | **S(Gev)** | **σ(Pb)** | **S(Gev)** | **σ(Pb)** |
| $m_{\left(\tilde{\chi}_i^+,\tilde{\chi}_j^-\right)}$ $\rightarrow m_{(600,800)}$ $\rightarrow m_{11}$ | 2361.4 | $5.7271 \times 10^{-17}$ | 2375.2 | $5.877 \times 10^{-17}$ | 2372.7 | $5.9544 \times 10^{-17}$ |
| $m_{\left(\tilde{\chi}_i^+,\tilde{\chi}_j^-\right)}$ $\rightarrow m_{(600,900)}$ $\rightarrow m_{12}$ | 2517.1 | $6.7647 \times 10^{-17}$ | 2525.9 | $6.9440 \times 10^{-17}$ | 2529.6 | $7.0650 \times 10^{-17}$ |
| $m_{\left(\tilde{\chi}_i^+,\tilde{\chi}_j^-\right)}$ $\rightarrow m_{(700,800)}$ $\rightarrow m_{21}$ | 2513.3 | $7.7699 \times 10^{-17}$ | 2520.8 | $7.9642 \times 10^{-17}$ | 2532.1 | $8.1103 \times 10^{-17}$ |
| $m_{\left(\tilde{\chi}_i^+,\tilde{\chi}_j^-\right)}$ $\rightarrow m_{(700,900)}$ $\rightarrow m_{22}$ | 2597.4 | $8.5805 \times 10^{-17}$ | 2599.9 | $8.6577 \times 10^{-17}$ | 2597.4 | $8.6003 \times 10^{-17}$ |

By investigation and by the Feynman rules, we computed the cross sections (σ) as a function of center of mass energy (S) for the process $e^-(p_1) + e^+(p_3) \rightarrow \tilde{\chi}_i^+(p_2) + \tilde{\chi}_j^-(p_4) + H_\ell^o(p_5)$ via $h^o$ and $Z^o$ propagators. Figs.2 (a-c) show that, as S increase from 1600 to 4000, a maximum values for the cross-sections are diverge at varies values of Chargino mass ($m_{\tilde{\chi}_i^+}, m_{\tilde{\chi}_j^-}$) and different value of neutral Higgs boson mass( $M_{h^o}, M_{A^o}, M_{H^o}$ ). From table (1) the best value of σ is ($8.6577 \times 10^{-17}$) Pb when masses of Charginos are $m_{\tilde{\chi}_i^-} = 900$ GeV, $m_{\tilde{\chi}_j^+} = 700$ GeV and $m_{H_\ell^o} = 135$ GeV

## 3.2  Group (II) via $Z^o$ and $h^o$ are the propagators

### 3.2.1 Feynman Diagram for Group (II)

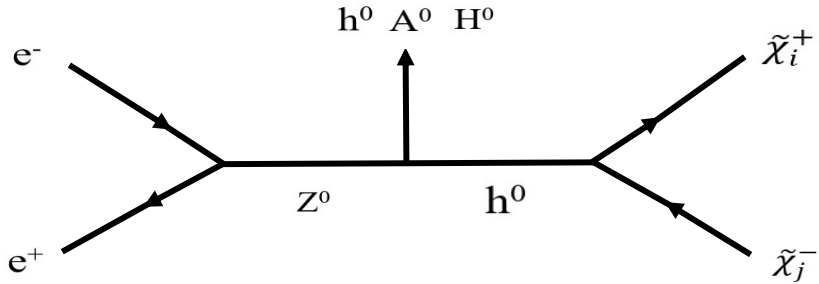

**Fig. (3):** Feynman diagram for the process $e^-(p_1) + e^+(p_3) \rightarrow \tilde{\chi}_i^+(p_2) + \tilde{\chi}_j^-(p_4) + H_\ell^o(p_5)$ via $Z^o$ and $h^o$

### 3.2.2 Matrix element and probability of reaction for Group (II)

**There should be 36 possibilities from (37 − 72) and the Matrix element is:**

$M_{(37-72)} = U_{e^-}(P_1)\bar{V}_{e^+}(P_3)\ (iG_m(p-q)^\mu)\ (\sigma^2 - m_H^2)^{-1}(S^2 - m_Z^2)^{-1}(\frac{-g}{2\cos\theta_w}\gamma_\alpha\ (\frac{1}{2} - 2Q_i\sin^2\theta_w - \frac{1}{2}\gamma_5))$

$V_{\tilde{\chi}^+}(P_2)\bar{U}_{\tilde{\chi}^-}(P_4)\ (-i\Upsilon^{\phi 0 X j^+\ X i^-})$ [34,36]

Where:

$G_m = ig(\frac{Sin(\beta-\alpha)}{2\,Cos\,\theta_w})$

$g$ : The gauge coupling constants of SU(2)$_L$

$g = \frac{e}{Sin\,\theta_w} = 0.637132$ [35,36]

$\Upsilon^{\phi 0 X j^+\ X i^-} = \frac{g}{\sqrt{2}}\ (K^*_{u\phi^\circ}U^*_{i1}V^*_{j2} + K^*_{d\phi^\circ}U^*_{i2}V^*_{j1})$

$K^*_{u\phi^\circ} = (cos\,\alpha\ ,\,sin\,\alpha\ ,\,i\,cos\,\beta_\circ\ ,\,i\,sin\,\beta_\circ),\,K^*_{d\phi^\circ} = (-sin\,\alpha\ ,\,cos\,\alpha\ ,\,i\,sin\,\beta_\circ\ ,\,-i\,cos\,\beta_\circ)$

 U and V are unitary matrices

*As, $\beta = 56.3,\ \alpha = -34.48$*

*$Q_i$ is a unitary matrix*

### 3.2.3 Cross Section Calculations in (Pb) for Group (II):

The cross sections as a function of center of mass energy for the Feynman diagrams of fig. (3) calculated and the results given in fig.4 (a-c) by interchanging the mass of Charginos ($m_{\tilde{\chi}_j^+}$, $m_{\tilde{\chi}_1^-}$) and the mass of neutral

Higgs boson ( $M_{h^0}$, $M_{A^0}$, $M_{H^0}$)

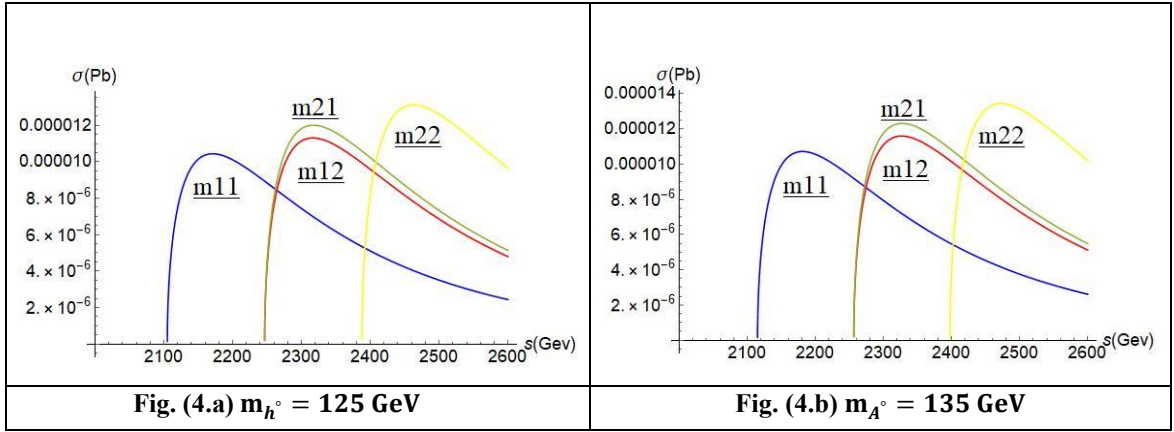

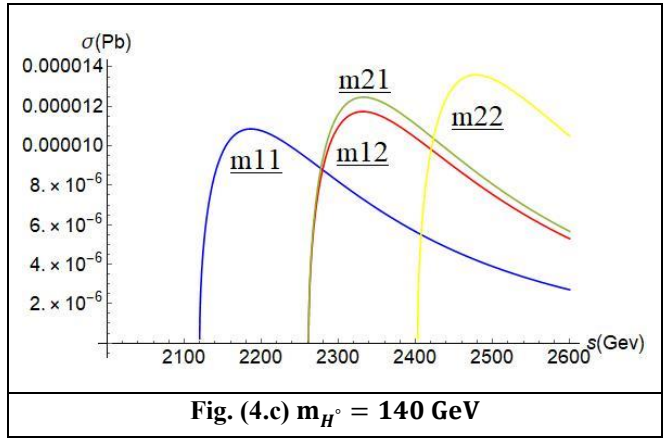

**Fig.4 (a-c):** The cross sections for the process $e^-(p_1) + e^+(p_3) \rightarrow \tilde{\chi}_i^+(p_2) + \tilde{\chi}_j^-(p_4) + H_\ell^o(p_5)$ as a function of center of mass energy via $Z^o$ and $h^o$ propagators by interchanging the mass of charginos $(m_{\tilde{\chi}_i^+}, m_{\tilde{\chi}_j^-})$ at different mass of neutral Higgs boson ( $M_{h^o}, M_{A^o}, M_{H^o}$ )

### 3.2.4 Comparing Results and Discussion for Group (II):

**Table (2):** Comparing between the cross sections for the process $e^-(P_1) + e^+(P_3) \rightarrow Z^o(P_1 + P_3) \rightarrow h^o(P_2 + P_4) \rightarrow \tilde{\chi}_i^+(p_2) + \tilde{\chi}_j^-(p_4)$ via $Z^o$ and $h^o$ propagators by interchanging the masses of Charginos ($m_{\tilde{\chi}_i^+}, m_{\tilde{\chi}_j^-}$) at different mass of resultant neutral Higgs boson ( $M_{h^o}, M_{A^o}, M_{H^o}$)

| $e^-(P_1) + e^+(P_3) \rightarrow Z^o(P_1 + P_3) \rightarrow h^o(P_2 + P_4) \rightarrow \tilde{\chi}_i^+(p_2) + \tilde{\chi}_j^-(p_4)$ | | | | | | |
|---|---|---|---|---|---|---|
| $m_{\tilde{\chi}_i^+}, m_{\tilde{\chi}_j^-}$ i,j = 1,2 | Resultant $m_{h^o}$ =125 | | Resultant $m_{A^o}$=135 | | Resultant $m_{H^o}$=140 | |
| | Fig. (4.a) | | Fig. (4.b) | | Fig. (4.c) | |
| | S(Gev) | $\sigma$(Pb) | S(Gev) | $\sigma$(Pb) | S(Gev) | $\sigma$(Pb) |
| $m_{\left(\tilde{\chi}_i^+ , \tilde{\chi}_j^-\right)}$ $\rightarrow m_{(600 , 800)}$ $\rightarrow m_{11}$ | 2167.8 | $1.0348 \times 10^{-5}$ | 2180.3 | $1.0651 \times 10^{-5}$ | 2184.1 | $1.0827 \times 10^{-5}$ |
| $m_{\left(\tilde{\chi}_i^+ , \tilde{\chi}_j^-\right)}$ $\rightarrow m_{(600 , 900)}$ $\rightarrow m_{12}$ | 2315.6 | $1.1246 \times 10^{-5}$ | 2325.6 | $1.1571 \times 10^{-5}$ | 2331.9 | $1.1709 \times 10^{-5}$ |
| $m_{\left(\tilde{\chi}_i^+ , \tilde{\chi}_j^-\right)}$ $\rightarrow m_{(700 , 800)}$ $\rightarrow m_{21}$ | 2315.6 | $1.1954 \times 10^{-5}$ | 2325.6 | $1.2249 \times 10^{-5}$ | 2331.9 | $1.2395 \times 10^{-5}$ |
| $m_{\left(\tilde{\chi}_i^+ , \tilde{\chi}_j^-\right)}$ $\rightarrow m_{(700 , 900)}$ $\rightarrow m_{22}$ | 2463.4 | $1.3089 \times 10^{-5}$ | 2470.9 | $1.3363 \times 10^{-5}$ | 2477.2 | $1.3572 \times 10^{-5}$ |

By investigation and by the Feynman rules, we computed the cross sections ($\sigma$) as a function of center of mass energy (S) for the process $e^-(p_1) + e^+(p_3) \rightarrow \tilde{\chi}_i^+(p_2) + \tilde{\chi}_j^-(p_4) + H_\ell^o(p_5)$ via $Z^o$ and $h^o$ propagators. Figs.4 (a-c) show that, as S increase from 1600 to 4000, a maximum values for the cross-sections is diverge at varies values of Chargino mass ($m_{\tilde{\chi}_i^+}, m_{\tilde{\chi}_j^-}$) and different value of neutral Higgs boson mass( $M_{h^o}, M_{A^o}, M_{H^o}$). From table (2) the ultimate value of $\sigma$ is $\left(1.3572 \times 10^{-5}\right)$ Pb when masses of Charginos are $m_{\tilde{\chi}_i^-} = 900$ GeV, $m_{\tilde{\chi}_j^+} = 700$ GeV and $m_{H_\ell^o} = 140$ GeV

## 3.3 Group (III) via $h^o$ and $h^o$ are the propagators

### 3.3.1 Feynman Diagram for Group (III)

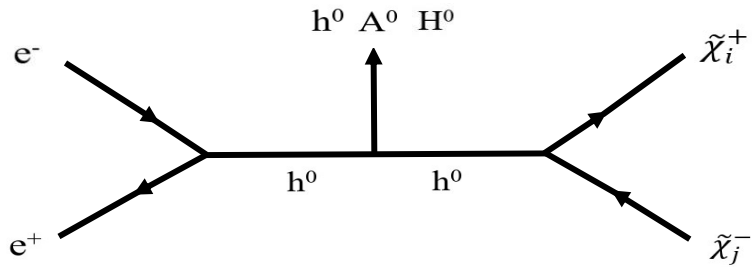

**Fig. (5):** Feynman diagram for the process $e^-(p_1) + e^+(p_3) \rightarrow \tilde{\chi}_i^+(p_2) + \tilde{\chi}_j^-(p_4) + H_\ell^o(p_5)$ via $h^o$ and $h^o$ there are (73-108) diagrams.

### 3.3.2   Matrix element and Probability of reaction for Group (III)

**There should be 36 possibilities from (73 – 109) and the Matrix element is:**

$$M_{(37-72)} = (-i(\ F_L + F_L^{'}\ \gamma_5)U_{e^-}(P_1)\bar{V}_{e^+}(P_3)\left(\frac{-3g\ m_H^2}{2m_z\ cos\ \theta_w}\right)(\sigma^2 - m_H^2)^{-1}(S^2 - m_H^2)^{-1}\left(\frac{-g\ m_H^2}{2m_z\ cos\ \theta_w}\ g^{\alpha\beta}\right)$$

$$V_{\tilde{\chi}^+}(P_2)\bar{U}_{\tilde{\chi}^-}(P_4)$$

*Where:*

$$F_L = -g\left(\frac{m_e Cos\alpha}{2M_W\ Cos\ \beta}\right),\ F_L^{'} = g\left(\frac{m_e Cos\alpha}{2M_W\ Cos\ \beta}\right)$$

$g^{\alpha\beta}$ is a (symmetric 4 x 4) metric tensor

### 3.3.3   Calculation Cross Sections in (Pb) for Group (III):

The Cross sections as a function of center of mass energy for the Feynman diagrams of fig. (5) have been calculated and the results are shown in fig.6 (a-c) by interchanging the mass of ($m_{\tilde{\chi}_j^+}$, $m_{\tilde{\chi}_1^-}$) and the mass of Neutral Higgs boson ( $M_{h^o}$, $M_{A^o}$, $M_{H^o}$)

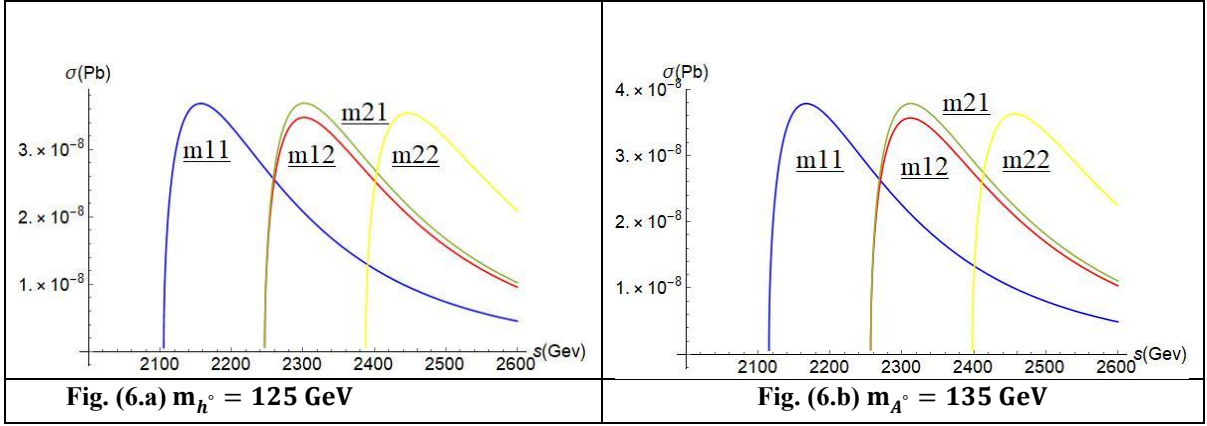

**Fig. (6.a) $m_{h^\circ}$ = 125 GeV** | **Fig. (6.b) $m_{A^\circ}$ = 135 GeV**

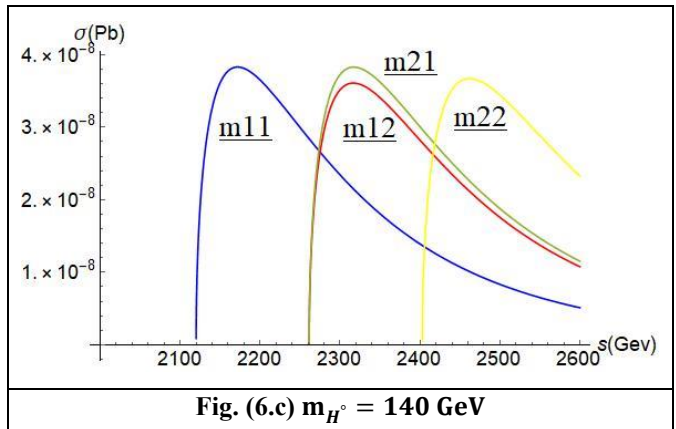

**Fig. (6.c) $m_{H^\circ}$ = 140 GeV**

**Fig.6(a-c):** The cross sections for the process $e^-(p_1) + e^+(p_3) \rightarrow \tilde{\chi}_t^+(p_2) + \tilde{\chi}_j^-(p_4) + H_e^*(p_5)$ as a function of center of mass energy via $h^o$ and $h^o$ propagators by interchanging the mass of ($m_{\tilde{\chi}_j^+}$, $m_{\tilde{\chi}_1^-}$) and the mass of neutral Higgs boson ( $M_{h^o}$, $M_{A^o}$, $M_{H^o}$) .

### 3.3.4 Comparing Results and Discussion for Group (III):

**Table(3):** comparing cross sections for the process $e^-(P_1) + e^+(P_3) \rightarrow h^o(P_1 + P_3) \rightarrow h^o(P_2 + P_4) \rightarrow \tilde{\chi}_i^+(p_2) + \tilde{\chi}_j^-(p_4)$ via $h^o$ and $h^o$ Propagators by interchanging the mass of $(m_{\tilde{\chi}_i^+}, m_{\tilde{\chi}_j^-})$ and the mass of neutral Higgs boson ( $M_{h^o}, M_{A^o}, M_{H^o}$ )

| $e^-(P_1) + e^+(P_3) \rightarrow h^o(P_1 + P_3) \rightarrow h^o(P_2 + P_4) \rightarrow \tilde{\chi}_i^+(p_2) + \tilde{\chi}_j^-(p_4)$ | | | | | | |
|---|---|---|---|---|---|---|
| $m_{\tilde{\chi}_i^+}, m_{\tilde{\chi}_j^-}$ **i,j = 1,2** | **Resutant $m_{h^o}$ =125** | | **Resultant $m_{A^o}$=135** | | **Resutant $m_{H^o}$=140** | |
| | **Fig. (6.a)** | | **Fig. (6.b)** | | **Fig. (6.c)** | |
| | S(Gev) | σ(Pb) | S(Gev) | σ(Pb) | S(Gev) | σ(Pb) |
| $m_{(\tilde{\chi}_i^+ , \tilde{\chi}_j^-)}$ $\rightarrow m_{(600 , 800)}$ $\rightarrow m_{11}$ | 2155.3 | 3.6834 $\times 10^{-8}$ | 2165.2 | $3.7766 \times 10^{-8}$ | 2171.4 | $3.8104 \times 10^{-8}$ |
| $m_{(\tilde{\chi}_i^+ , \tilde{\chi}_j^-)}$ $\rightarrow m_{(600 , 900)}$ $\rightarrow m_{12}$ | 2300.6 | 3.4724 $\times 10^{-8}$ | 2311.8 | $3.5603 \times 10^{-8}$ | 2316.8 | $3.5914 \times 10^{-8}$ |
| $m_{(\tilde{\chi}_i^+ , \tilde{\chi}_j^-)}$ $\rightarrow m_{(700 , 800)}$ $\rightarrow m_{21}$ | 2301.9 | 3.6702 $\times 10^{-8}$ | 2313.0 | $3.7631 \times 10^{-8}$ | 2315.5 | $3.8241 \times 10^{-8}$ |
| $m_{(\tilde{\chi}_i^+ , \tilde{\chi}_j^-)}$ $\rightarrow m_{(700 , 900)}$ $\rightarrow m_{22}$ | 2447.2 | 3.5383 $\times 10^{-8}$ | 2457.2 | $3.6144 \times 10^{-8}$ | 2463.4 | $3.6461 \times 10^{-8}$ |

By investigation and by the Feynman rules, we computed the cross sections (σ) as a function of center of mass energy (S) for the process $e^-(p_1) + e^+(p_3) \rightarrow \tilde{\chi}_i^+(p_2) + \tilde{\chi}_j^-(p_4) + H_\ell^o(p_5)$ via $h^o$ and $h^o$ propagators. Figs.6 (a-c) show that, at S increase from 1600 to 3500, a maximum values for the cross-sections is diverge at varies values of Chargino mass ( $m_{\tilde{\chi}_i^+}, m_{\tilde{\chi}_j^-}$ ) and different value of neutral Higgs boson mass( $M_{h^o}, M_{A^o}, M_{H^o}$ ). From table (3) the best value of σ is $(3.8241 \times 10^{-8})$ Pb when masses of Charginos are $m_{\tilde{\chi}_1^-} = 800$ GeV, $m_{\tilde{\chi}_j^+} = 700$ GeV and $m_{H_\ell^o} = 140$ GeV

## 3.4 Group (IV) $Z^o$ and $Z^o$ are the propagators

### 3.4.1 Feynman Diagram for Group (IV)

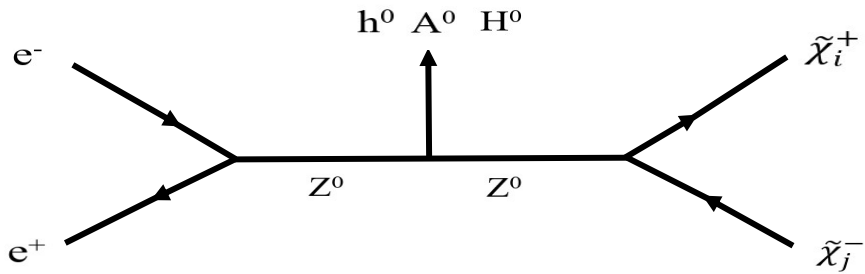

**Fig. (7):** Feynman diagram for the process $e^-(p_1) + e^+(p_3) \rightarrow \tilde{\chi}_i^+(p_2) + \tilde{\chi}_j^-(p_4) + H_\ell^o(p_5)$ via $Z^o$ and $Z^o$ there are (109-144) diagrams.

### 3.4.2 Matrix element and Probability of reaction for Group (IV)

**There should be 36 possibilities from (109 – 144) and the Matrix element is:**

$$M_{(109-144)} = (\frac{-g}{2\cos\theta_w})\gamma_\alpha \ (\frac{1}{2} - 2Qi\sin^2\theta_w - \frac{1}{2}\gamma_5))U_{e^-}(P_1)\bar{V}_{e^+}(P_3) \ (ig(\frac{m_Z\cos(\beta-\alpha)}{\cos\theta_w})g^{\mu\nu})(\sigma^2 -$$

$$m_Z^2)^{-1}(S^2 - m_Z^2)^{-1}(e\frac{\cos^2\theta_w - \sin^2\theta_w}{2\sin\theta_w\cos\theta_w}(P_2 - P_1)) \ V_{\tilde{\chi}^+}(P_2)\bar{U}_{\tilde{\chi}^-}(P_4) \ [36,37]$$

Where:

$Q_i$ *is a unitary matrix ,* $g^{\mu\nu}$ is a (symmetric 4 x 4) metric tensor

*As,* $\beta$ = 56.3, $\alpha$ = - 34.48, $\theta_w$ =28.7, e = 0.302822

### 3.4.3 Cross Section Calculations in (Pb) for Group (IV):

The cross sections as a function of center of mass energy for the Feynman diagrams of fig. (7) have been calculated and the results are shown in fig.8 (a-c) by interchanging the mass of charginos ($m_{\tilde{\chi}_j^+}$, $m_{\tilde{\chi}_1^-}$) and the mass of neutral Higgs boson ( $M_{h^0}, M_{A^0}, M_{H^0}$)

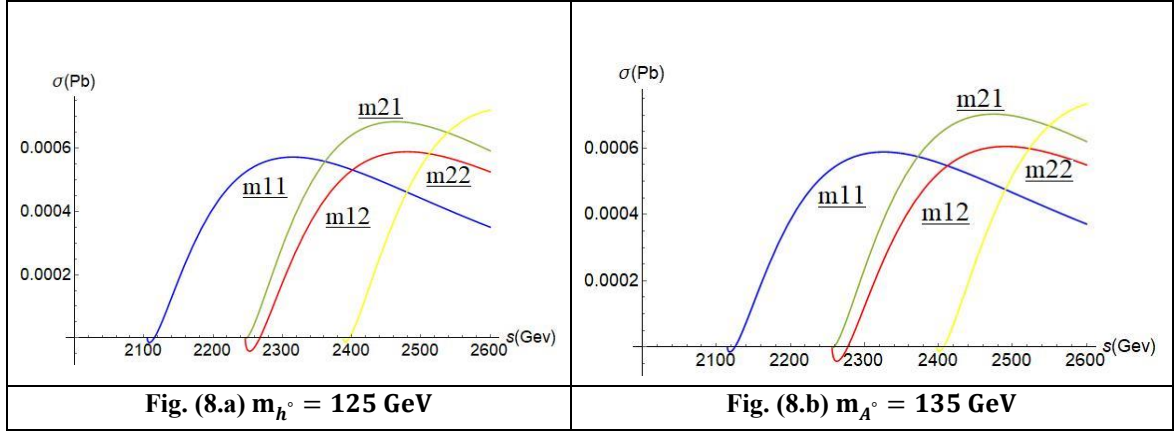

**Fig. (8.a) $m_{h^\circ}$ = 125 GeV**    **Fig. (8.b) $m_{A^\circ}$ = 135 GeV**

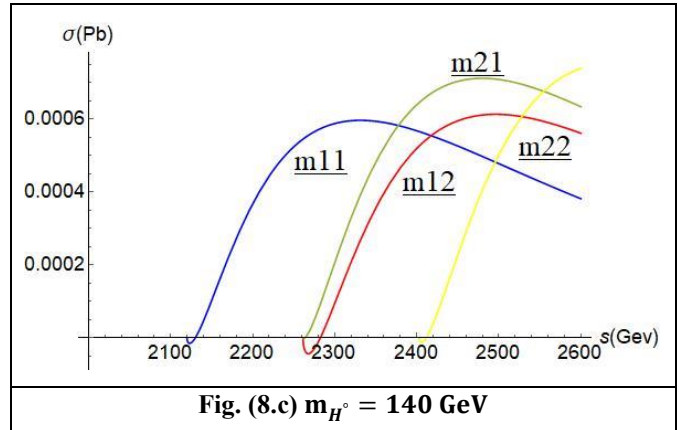

**Fig. (8.c) $m_{H^\circ}$ = 140 GeV**

**Fig.8 (a-c):** The cross sections for the process $e^-(p_1) + e^+(p_3) \rightarrow \tilde{\chi}_i^+(p_2) + \tilde{\chi}_j^-(p_4) + H_e^\circ(p_5)$ as a function of center of mass energy via $Z^o$ and $Z^o$ propagators by interchanging the mass of ($m_{\tilde{\chi}_j^+}$, $m_{\tilde{\chi}_1^-}$) and the mass of neutral Higgs boson ( $M_{h^0}, M_{A^0}, M_{H^0}$).

### 3.4.4    Comparing Results and Discussion for Group (IV):

**Table(4):** cross sections for the process $e^-(P_1) + e^+(P_3) \rightarrow Z^o(P_1 + P_3) \rightarrow Z^o(P_2 + P_4) \rightarrow \tilde{\chi}_i^+(p_2) + \tilde{\chi}_j^-(p_4)$ via $Z^o$ and $Z^o$ Propagators by interchanging the mass of $(m_{\tilde{\chi}_j^+}, m_{\tilde{\chi}_i^-})$ and the mass of neutral Higgs boson ( $M_{h^o}, M_{A^o}, M_{H^o}$ )

| $e^-(P_1) + e^+(P_3) \rightarrow Z^o(P_1 + P_3) \rightarrow Z^o(P_2 + P_4) \rightarrow \tilde{\chi}_i^+(p_2) + \tilde{\chi}_j^-(p_4)$ | | | | | | |
|---|---|---|---|---|---|---|
| $m_{\tilde{\chi}_i^+}, m_{\tilde{\chi}_j^-}$ $i,j = 1,2$ | **Resultant $m_{h^o}$ =125** | | **Resultant $m_{A^o}$=135** | | **Resultant $m_{H^o}$=140** | |
| | **Fig. (8.a)** | | **Fig. (8.b)** | | **Fig. (8.c)** | |
| | S(Gev) | σ(Pb) | S(Gev) | σ(Pb) | S(Gev) | σ(Pb) |
| $m_{\left(\tilde{\chi}_i^+, \tilde{\chi}_j^-\right)}$ $\rightarrow m_{(600,800)}$ $\rightarrow m_{11}$ | 2303.1 | $5.6681 \times 10^{-4}$ | 2316.6 | $5.8589 \times 10^{-4}$ | 2323.9 | $5.9607 \times 10^{-4}$ |
| $m_{\left(\tilde{\chi}_i^+, \tilde{\chi}_j^-\right)}$ $\rightarrow m_{(600,900)}$ $\rightarrow m_{12}$ | 2465.7 | $5.8557 \times 10^{-4}$ | 2478.0 | $5.9955 \times 10^{-4}$ | 2480.4 | $6.0985 \times 10^{-4}$ |
| $m_{\left(\tilde{\chi}_i^+, \tilde{\chi}_j^-\right)}$ $\rightarrow m_{(700,800)}$ $\rightarrow m_{21}$ | 2457.2 | $6.8206 \times 10^{-4}$ | 2470.6 | $7.0339 \times 10^{-4}$ | 2475.5 | $7.1179 \times 10^{-4}$ |
| $m_{\left(\tilde{\chi}_i^+, \tilde{\chi}_j^-\right)}$ $\rightarrow m_{(700,900)}$ $\rightarrow m_{22}$ | 2596.6 | $7.1690 \times 10^{-4}$ | 2600.2 | $7.3345 \times 10^{-4}$ | 2600.2 | $7.3934 \times 10^{-4}$ |

By investigation and by the Feynman rules, we computed the cross sections (σ) as a function of center of mass energy (S) for the process $e^-(p_1) + e^+(p_3) \rightarrow \tilde{\chi}_i^+(p_2) + \tilde{\chi}_j^-(p_4) + H_\ell^o(p_5)$ via $Z^o$ and $Z^o$ propagators. Figs.8 (a-c) show that, as S increase from 1600 to 3500, a maximum values for the cross-sections is diverge at varies values of Chargino mass $(m_{\tilde{\chi}_i^+}, m_{\tilde{\chi}_j^-})$ and different value of neutral Higgs boson mass( $M_{h^o}, M_{A^o}, M_{H^o}$ ). From table (4) the best value of σ is $\left(7.3934 \times 10^{-4}\right)$ Pb when masses of Charginos are $m_{\tilde{\chi}_i^-} = 900$ GeV, $m_{\tilde{\chi}_j^+} = 700$ GeV and $m_{H_\ell^o} = 140$ GeV

### 3.5 Group (V) via $\widetilde{\chi}^o$ and $Z^o$ are the propagators

#### 3.5.1 Feynman Diagram for Group (V)

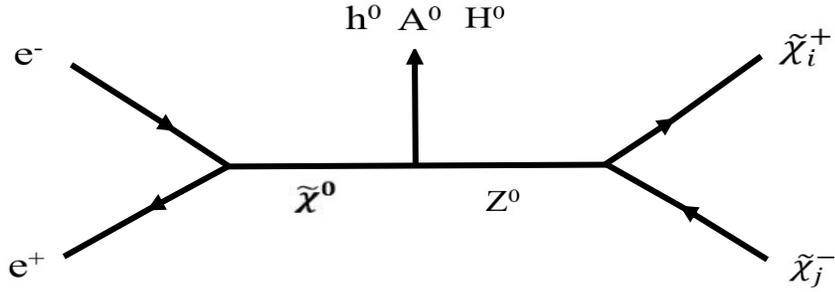

**Fig. (9):** Feynman diagram for the process $e^-(p_1) + e^+(p_3) \rightarrow \widetilde{\chi}_i^+(p_2) + \widetilde{\chi}_j^-(p_4) + H_\ell^o(p_5)$ via $\widetilde{\chi}^o$ and $Z^o$ there are (145-180) diagrams.

#### 3.5.2 Matrix element and Probability of reaction for Group (V)

**There should be 36 possibilities from (145-180) and the Matrix element is:**

$M_{(145-180)} = (-i\,e\,\frac{1}{2\,Sin\,\theta_w}\frac{m_f}{m_w}\,\gamma_5)U_{e^-}(P_1)\bar{V}_{e^+}(P_3)\,(\frac{i\,e}{2\,Sin\,\theta_w Cos\,\theta_w}\,((1-\varepsilon)P_2 - (1+\varepsilon)P_4))\,(\sigma^2 -$

$m_Z^2)^{-1}(S^2 - m_{\widetilde{\chi}^o}^2)^{-1}\,(e\,\frac{\cos^2\theta_w - \sin^2\theta_w}{2\,\sin\theta_w\,\cos\theta_w}(P_2 - P_1))\,V_{\widetilde{\chi}^+}(P_2)\bar{U}_{\widetilde{\chi}^-}(P_4)$ [38,39]

Where:

$m_f$, is mass of electron = 0.000511 Gev/C$^2$ ,$m_w$ = 80 Gev/C$^2$

$\varepsilon_{ij}$ = 1, because we use different masses of charginos, e = 0.302822

#### 3.5.3 Cross Section Calculations in (Pb) for Group (V):

The cross sections as a function of center of mass energy for the Feynman diagrams of fig. (9) have been calculated and the results are shown in fig.10 (a-c) by interchanging the mass of charginos ($m_{\widetilde{\chi}_j^+}$, $m_{\widetilde{\chi}_i^-}$) and

the mass of neutral Higgs boson ( $M_{h^o}$,$M_{A^o}$,$M_{H^o}$)

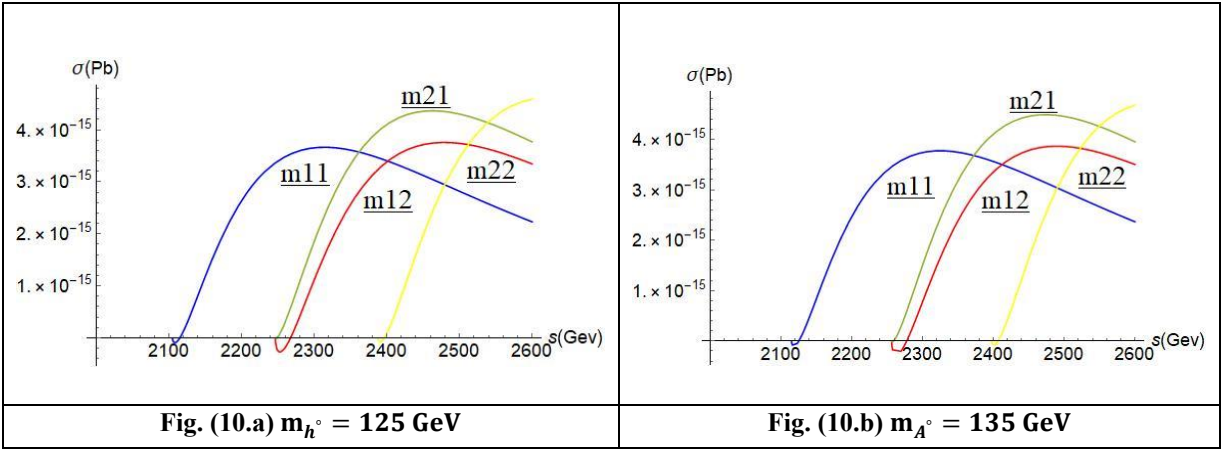

| Fig. (10.a) $m_{h^\circ} = 125$ GeV | Fig. (10.b) $m_{A^\circ} = 135$ GeV |

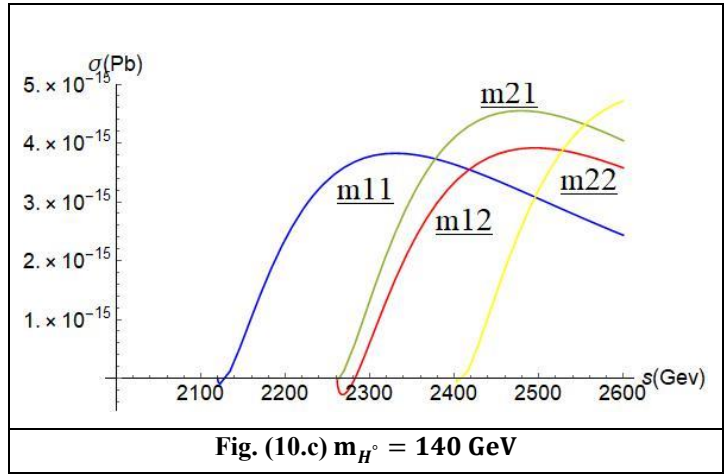

| Fig. (10.c) $m_{H^\circ} = 140$ GeV |

**Fig.10(a-c):** The cross sections for the process $e^-(p_1) + e^+(p_3) \rightarrow \tilde{\chi}_i^+(p_2) + \tilde{\chi}_j^-(p_4) + H_\ell^\circ(p_5)$ as a function of center of mass energy via $\tilde{\chi}^o$ and $Z^o$ propagators by interchanging the mass of $(m_{\tilde{\chi}_j^+}, m_{\tilde{\chi}_1^-})$ and the mass of neutral Higgs boson $(M_{h^o}, M_{A^o}, M_{H^o})$.

### 3.5.4   Comparing Results and Discussion for Group (V):

**Table(5):**  cross sections for the process $e^-(P_1) + e^+(P_3) \rightarrow \tilde{\chi}^o(P_1 + P_3) \rightarrow Z^o(P_2 + P_4) \rightarrow \tilde{\chi}_i^+(p_2) + \tilde{\chi}_j^-(p_4)$ via $\tilde{\chi}^o$ and $Z^o$ Propagators by interchanging the mass of $(m_{\tilde{\chi}_j^+}, m_{\tilde{\chi}_i^-})$ and the mass of neutral Higgs boson ( $M_{h^o}, M_{A^o}, M_{H^o}$ ).

| $e^-(P_1) + e^+(P_3) \rightarrow \tilde{\chi}^o(P_1 + P_3) \rightarrow Z^o(P_2 + P_4) \rightarrow \tilde{\chi}_i^+(p_2) + \tilde{\chi}_j^-(p_4)$ | | | | | |
|---|---|---|---|---|---|
| $m_{\tilde{\chi}_i^+}, m_{\tilde{\chi}_i^-}$ $i,j = 1,2$ | **Resultant $m_{h^o}$ =125** | | **Resultant $m_{A^o}$ =135** | | **Resultant $m_{H^o}$ =140** |
| | **Fig. (10.a)** | | **Fig. (10.b)** | | **Fig. (10.c)** |
| | S(Gev) | σ(Pb) | S(Gev) | σ(Pb) | S(Gev) | σ(Pb) |
| $m_{(\tilde{\chi}_i^+ \cdot \tilde{\chi}_j^-)}$ $\rightarrow m_{(600 \cdot 800)}$ $\rightarrow m_{11}$ | 2299.9 | $3.6296 \times 10^{-15}$ | 2312.5 | $3.7490 \times 10^{-15}$ | 2322.5 | $3.8017 \times 10^{-15}$ |
| $m_{(\tilde{\chi}_i^+ \cdot \tilde{\chi}_j^-)}$ $\rightarrow m_{(600 \cdot 900)}$ $\rightarrow m_{12}$ | 2466.9 | $3.7348 \times 10^{-15}$ | 2475.6 | $3.8371 \times 10^{-15}$ | 2485.7 | $3.8918 \times 10^{-15}$ |
| $m_{(\tilde{\chi}_i^+ \cdot \tilde{\chi}_j^-)}$ $\rightarrow m_{(700 \cdot 800)}$ $\rightarrow m_{21}$ | 2453.1 | $4.3486 \times 10^{-15}$ | 2470.6 | $4.4712 \times 10^{-15}$ | 2475.6 | $4.5401 \times 10^{-15}$ |
| $m_{(\tilde{\chi}_i^+ \cdot \tilde{\chi}_j^-)}$ $\rightarrow m_{(700 \cdot 900)}$ $\rightarrow m_{22}$ | 2596.1 | $4.5590 \times 10^{-15}$ | 2598.6 | $4.6473 \times 10^{-15}$ | 2598.6 | $4.6842 \times 10^{-15}$ |

By investigation and by the Feynman rules, we computed the cross sections (σ) as a function of center of mass energy (S) for the process $e^-(p_1) + e^+(p_3) \rightarrow \tilde{\chi}_i^+(p_2) + \tilde{\chi}_j^-(p_4) + H_\ell^o(p_5)$ via $\tilde{\chi}^o$ and $Z^o$ propagators. Figs.10 (a-c) show that, as S increase from 1600 to 3500, a maximum values for the cross-sections is diverge at varies values of Chargino mass ( $m_{\tilde{\chi}_i^+}, m_{\tilde{\chi}_j^-}$ ) and different value of neutral Higgs boson mass( $M_{h^o}, M_{A^o}, M_{H^o}$ ). From table (5) the best value of σ is ($4.6842 \times 10^{-15}$) Pb when masses of Charginos are  $m_{\tilde{\chi}_i^-} = 900$ GeV, $m_{\tilde{\chi}_j^+} = 700$ GeV  and $m_{H_\ell^o} = 140$ GeV

## 4.   Results and discussion

Based on the investigation of the production of neutral Higgs boson and two charged charginos from electron – positron annihilation via different propagators for interaction $e^-(p_1) + e^+(p_3) \rightarrow \tilde{\chi}_i^+(p_2) + \tilde{\chi}_j^-(p_4) + H_\ell^o(p_5)$  we can study the cross-sections σ (Pb) as a function of center of mass energy S (GeV). We estimated the cross sections for this interaction in the Minimal Supersymmetric Standard Model (MSSM).

We draw all Feynman diagrams probabilities for the reaction $e^-(p_1) + e^+(p_3) \rightarrow \tilde{\chi}_i^+(p_2) + \tilde{\chi}_j^-(p_4) + H_\ell^o(p_5)$ arranged according to the propagators, we have 180 probabilities of the reaction divided into five groups.

Groupe (I) Production via  $h^0$ and $Z^0$ propagators,

 $e^-(P_1) + e^+(P_3) \rightarrow h^0(P_1 + P_3) \rightarrow Z^o(P_2 + P_4) \rightarrow \tilde{\chi}_i^+(p_2) + \tilde{\chi}_j^-(p_4)$

Groupe (II) Production via $Z^0$ and $h^0$ propagators,

$$e^-(P_1) + e^+(P_3) \rightarrow Z^0(P_1 + P_3) \rightarrow h^o(P_2 + P_4) \rightarrow \tilde{\chi}_i^+(p_2) + \tilde{\chi}_j^-(p_4)$$

Groupe (III) Production via $h^0$ and $h^0$ propagators,

$$e^-(P_1) + e^+(P_3) \rightarrow h^0(P_1 + P_3) \rightarrow h^o(P_2 + P_4) \rightarrow \tilde{\chi}_i^+(p_2) + \tilde{\chi}_j^-(p_4)$$

Groupe (IV) Production via $Z^0$ and $Z^0$ propagators,

$$e^-(P_1) + e^+(P_3) \rightarrow Z^0(P_1 + P_3) \rightarrow Z^o(P_2 + P_4) \rightarrow \tilde{\chi}_i^+(p_2) + \tilde{\chi}_j^-(p_4)$$

Groupe (V) Production via $h^0$ and $Z^0$ propagators,

$$e^-(P_1) + e^+(P_3) \rightarrow \tilde{\chi}^o(P_1 + P_3) \rightarrow Z^0(P_2 + P_4) \rightarrow \tilde{\chi}_i^+(p_2) + \tilde{\chi}_j^-(p_4)$$

We calculated the cross – sections $\sigma$ (Pb) as a function of the center of mass energy S (GeV) at different values of Chargino mass $(m_{\tilde{\chi}_i^+}, m_{\tilde{\chi}_j^-})$ where, $m_{(\tilde{\chi}_i^+, \tilde{\chi}_j^-)} \rightarrow m_{(600,800)}$, $m_{(\tilde{\chi}_i^+, \tilde{\chi}_j^-)} \rightarrow m_{(600,900)}$,

$m_{(\tilde{\chi}_i^+, \tilde{\chi}_j^-)} \rightarrow m_{(700,800)} m_{(\tilde{\chi}_i^+, \tilde{\chi}_j^-)} \rightarrow m_{(700,900)}$, and different value of neutral Higgs boson mass ( $M_{h^0} = 125\ GeV$, $M_{A^0} = 135\ GeV$, $M_{H^0} = 140$ GeV) for fives groups.

And draw the cross—sections as a function of incident energy for all probabilities, there are 60 curves.

**By comparing the results of the peak values for cross- sections all curves we found that:**

**Table (6):** The peak values of the cross sections of the interaction $e^-(p_1) + e^+(p_3) \rightarrow \tilde{\chi}_i^+(p_2) + \tilde{\chi}_j^-(p_4) + H_\ell^{\circ}(p_5)$ via different propagators, with different masses of Charginos $(m_{\tilde{\chi}_i^+}, m_{\tilde{\chi}_j^-})$ and neutral Higgs boson mass $M_{H^0}$ at different values of incident energies for five groups.

| Group N$\Omega$ | $e^-(p_1) + e^+(p_3) \rightarrow \tilde{\chi}_i^+(p_2) + \tilde{\chi}_j^-(p_4) + H_\ell^{\circ}(p_5)$ | Fig. no. | $m_{h^0,A^0,H^0}$ (GeV) | $m_{\tilde{\chi}_i^+} m_{\tilde{\chi}_j^-}$ (GeV/ C$^2$) | S (GeV) | $\sigma_{max}$ (Pb) |
|---|---|---|---|---|---|---|
| 1$^{st}$ | Production via $h^0$ and $Z^0$ $e^-(P_1) + e^+(P_3) \rightarrow h^0(P_1 + P_3)$ $\rightarrow Z^0(P_2 + P_4) \rightarrow \tilde{\chi}_i^+(p_2) + \tilde{\chi}_j^-(p_4)$ | 2.b | $m_{H^0}$ $=135 GeV$ | 700,900 | 2599.9 | $8.6577 \times 10^{-17}$ |
| 2$^{nd}$ | Production via $Z^0$ and $h^0$ $e^-(P_1) + e^+(P_3) \rightarrow Z^o(P_1 + P_3)$ $\rightarrow h^o(P_2 + P_4) \rightarrow \tilde{\chi}_i^+(p_2) + \tilde{\chi}_j^-(p_4)$ | 4.c | $m_{H^0}$ $=140 GeV$ | 700,900 | 2477.2 | $1.3572 \times 10^{-5}$ |
| 3$^{rd}$ | Production via $h^0$ and $h^0$ $e^-(P_1) + e^+(P_3) \rightarrow h^o(P_1 + P_3)$ $\rightarrow h^o(P_2 + P_4) \rightarrow \tilde{\chi}_i^+(p_2) + \tilde{\chi}_j^-(p_4)$ | 6.c | $m_{H^0}$ $=140 GeV$ | 700,800 | 2315.5 | $3.8241 \times 10^{-8}$ |
| 4$^{th}$ | Production via $Z^0$ and $Z^0$ $e^-(P_1) + e^+(P_3) \rightarrow Z^o(P_1 + P_3)$ $\rightarrow Z^o(P_2 + P_4) \rightarrow \tilde{\chi}_i^+(p_2) + \tilde{\chi}_j^-(p_4)$ | 8.c | $m_{H^0}$ $=140 GeV$ | 700,900 | 2600.2 | $7.3934 \times 10^{-4}$ |
| 5$^{th}$ | Production via $\tilde{\chi}^o$ and $Z^0$ $e^-(P_1) + e^+(P_3) \rightarrow \tilde{\chi}^o(P_1 + P_3)$ $\rightarrow Z^o(P_2 + P_4) \rightarrow \tilde{\chi}_i^+(p_2) + \tilde{\chi}_j^-(p_4)$ | 10.c | $m_{H^0}$ $=140 GeV$ | 700,900 | 2598.6 | $4.6842 \times 10^{-15}$ |

**The dominant process**

Is Group (IV), $e^-(P_1) + e^+(P_3) \rightarrow Z^o(P_1 + P_3) \rightarrow Z^o(P_2 + P_4) \rightarrow \tilde{\chi}_i^+(p_2) + \tilde{\chi}_j^-(p_4)$

in which S interval (1600- 3500) Gev, the best value of $\sigma$ is ( $7.3934 \times 10^{-4}$ ) Pb. When masses of

Charginos are $m_{\tilde{\chi}_i^-} = 900$ GeV, $m_{\tilde{\chi}_j^+} = 700$ GeV and mass of neutral Higgs boson is $m_{H_\ell^\circ} = 140$ GeV

**The competing process**

Is Group (II), $e^-(P_1) + e^+(P_3) \rightarrow Z^o(P_1 + P_3) \rightarrow h^o(P_2 + P_4) \rightarrow \tilde{\chi}_i^+(p_2) + \tilde{\chi}_j^-(p_4)$

in which S interval (1600- 4000) Gev, the best value of $\sigma$ is ( $1.3572 \times 10^{-5}$ ) Pb. When masses of

Charginos are $m_{\tilde{\chi}_i^-} = 900$ GeV, $m_{\tilde{\chi}_j^+} = 700$ GeV and mass of neutral Higgs boson is $m_{H_\ell^\circ} = 140$ GeV

**And the other processes**

Are Group (III), $e^-(P_1) + e^+(P_3) \rightarrow h^o(P_1 + P_3) \rightarrow h^o(P_2 + P_4) \rightarrow \tilde{\chi}_i^+(p_2) + \tilde{\chi}_j^-(p_4)$

in which S interval (1600- 3500) Gev, the best value of $\sigma$ is ( $3.8241 \times 10^{-8}$ ) Pb. When masses of

Charginos are $m_{\tilde{\chi}_i^-} = 800$ GeV, $m_{\tilde{\chi}_j^+} = 700$GeV and mass of neutral Higgs boson is $m_{H_\ell^\circ} = 140$ GeV

Group (V), $e^-(P_1) + e^+(P_3) \rightarrow \tilde{\chi}^o(P_1 + P_3) \rightarrow Z^o(P_2 + P_4) \rightarrow \tilde{\chi}_i^+(p_2) + \tilde{\chi}_j^-(p_4)$

in which S interval (1600- 3500) Gev, the best value of $\sigma$ is ( $4.6842 \times 10^{-15}$ ) Pb. When masses of

Charginos are $m_{\tilde{\chi}_i^-} = 900$ GeV, $m_{\tilde{\chi}_j^+} = 700$GeV and mass of neutral Higgs boson is $m_{H_\ell^\circ} = 140$ GeV

Group (I), $e^-(P_1) + e^+(P_3) \rightarrow h^o(P_1 + P_3) \rightarrow Z^o(P_2 + P_4) \rightarrow \tilde{\chi}_i^+(p_2) + \tilde{\chi}_j^-(p_4)$

in which S interval (1600- 4000) Gev, the best value of $\sigma$ is ( $8.6577 \times 10^{-17}$ ) Pb. When masses of

Charginos are $m_{\tilde{\chi}_i^-} = 900$ GeV, $m_{\tilde{\chi}_j^+} = 700$GeV and mass of neutral Higgs boson is $m_{H_\ell^\circ} = 135$ GeV

## 5. Conclusion

In our investigation, we were successful in identifying the scenario with the highest cross section for the reaction. $e^-(p_1) + e^+(p_3) \rightarrow \tilde{\chi}_i^+(p_2) + \tilde{\chi}_j^-(p_4) + H_\ell^\circ(p_5)$ . We calculated the cross sections for this interaction in the Minimal Supersymmetric Standard Model (MSSM)

-The best cross section increases to $(7.3934 \times 10^{-4}$ Pb) at (S= 2600.2 GeV) when masses of Charginos are $m_{\tilde{\chi}_i^-} = 900$ GeV, $m_{\tilde{\chi}_j^+} = 700$ GeV and mass of neutral Higgs boson is $m_{H_\ell^\circ} = 140$ GeV, via $Z^0$ and $Z^0$

boson propagators exchange in Fig. (3.c) for the reaction Production via $Z^0$ and $Z^0$

$e^-(P_1) + e^+(P_3) \rightarrow Z^o(P_1 + P_3) \rightarrow Z^o(P_2 + P_4) \rightarrow \tilde{\chi}_i^+(p_2) + \tilde{\chi}_j^-(p_4)$ in group (IV)